\documentclass[aps,pra,twocolumn]{revtex4-1}
\usepackage{amsmath}
\usepackage{amssymb}
\usepackage{epsfig}
\usepackage{graphics}
\usepackage{bm}

\begin{document}
\title{A long-lived Higgs  mode in a two-dimensional confined Fermi gas}
\author{G.\ M.\ Bruun}
\affiliation{Department of Physics and Astronomy, University of Aarhus, Ny Munkegade, DK-8000 Aarhus C, Denmark}

\begin{abstract}
The Higgs mode corresponds to the collective motion of particles due to the vibrations of an invisible field. It plays a fundamental role for our understanding of both low and high energy physics, giving elementary particles their mass and  leading to collective modes in 
 condensed matter  and  nuclear systems.  The Higgs mode has been observed in a limited number of table-top systems, where it however is characterised by a short lifetime 
 due to decay into a continuum of modes. 
  A major goal which has remained elusive so far, is therefore to realise  a long-lived Higgs mode in a controllable system. 
Here, we show how an undamped Higgs mode can be observed unambiguously in a  Fermi gas in a two-dimensional trap, 
 close to a quantum phase transition between a normal and a superfluid phase.  We develop a first-principles theory of the pairing  and the associated 
 collective  modes, which
 is  quantitatively reliable when the pairing energy is much smaller than the trap level spacing, yet simple enough to allow the derivation of  analytical results.
The theory includes the trapping potential exactly, which is demonstrated to stabilize the Higgs mode by making its decay channels discrete.  
 Our results show how atoms in micro-traps can unravel properties of a long-lived Higgs mode, 
including the role of confinement and finite size effects.
\end{abstract}
\maketitle

According to the standard description of many-body systems, the  broken symmetry state  of a system is characterised by a
collective field. Phase oscillations of this field gives rise to  massless Goldstone  modes, whereas  oscillations in the amplitude correspond to the 
massive Higgs  mode~\cite{Anderson1958,Bogoliubov1958,Higgs1964}. 
    The Higgs mode plays a fundamental role in our understanding of nature across many energy scales: It leads to the existence of collective modes 
    in condensed matter systems~\cite{Sachdev2011book}, pair vibration modes in atomic nuclei~\cite{BohrMottelson1975book,Potel2013}, and 
    in the Standard Model
    it gives elementary particles their mass ~\cite{Higgs1964}, as was recently confirmed in two spectacular experiments at CERN~\cite{CMS2012,ATLAS2012}.   It is therefore desirable  to have 
     controllable table-top systems where one can investigate fundamental questions such as the existence of a sharp Higgs mode
     which is stable against decay~\cite{Podolsky2011}, 
 and  the interplay between confinement  and the Higgs mode, which is relevant for atomic nuclei as well as 
 particle physics models with compact dimensions~\cite{Randall1999,Arkani-Hamed1998}. Also, since the Higgs mode is a consequence of the broken symmetry of 
 a many-particle state, an intriguing question is to examine how this mode emerges with increasing particle number. 
The list of table-top systems where the Higgs mode has been observed is fairly short. Raman scattering data for a 
 niobium selenide superconductor are consistent with a Higgs mode coupled to a
 charge-density wave~\cite{Measson2014,Littlewood1981,Sooryakumar1980}. Also, the Higgs mode has been identified in 
  neutron scattering experiments for a  quantum anti-ferromagnet~\cite{Ruegg2008}, and in lattice modulation 
 experiments for a gas of bosonic atoms in an optical lattice~\cite{Endres2012,Bissbort2011}. In these cases, the spectral signal of the Higgs mode 
 is however very broad due to decay into a continuum of  modes  which complicates a quantitative analysis.  
\begin{figure}
\includegraphics[width=0.7\columnwidth]{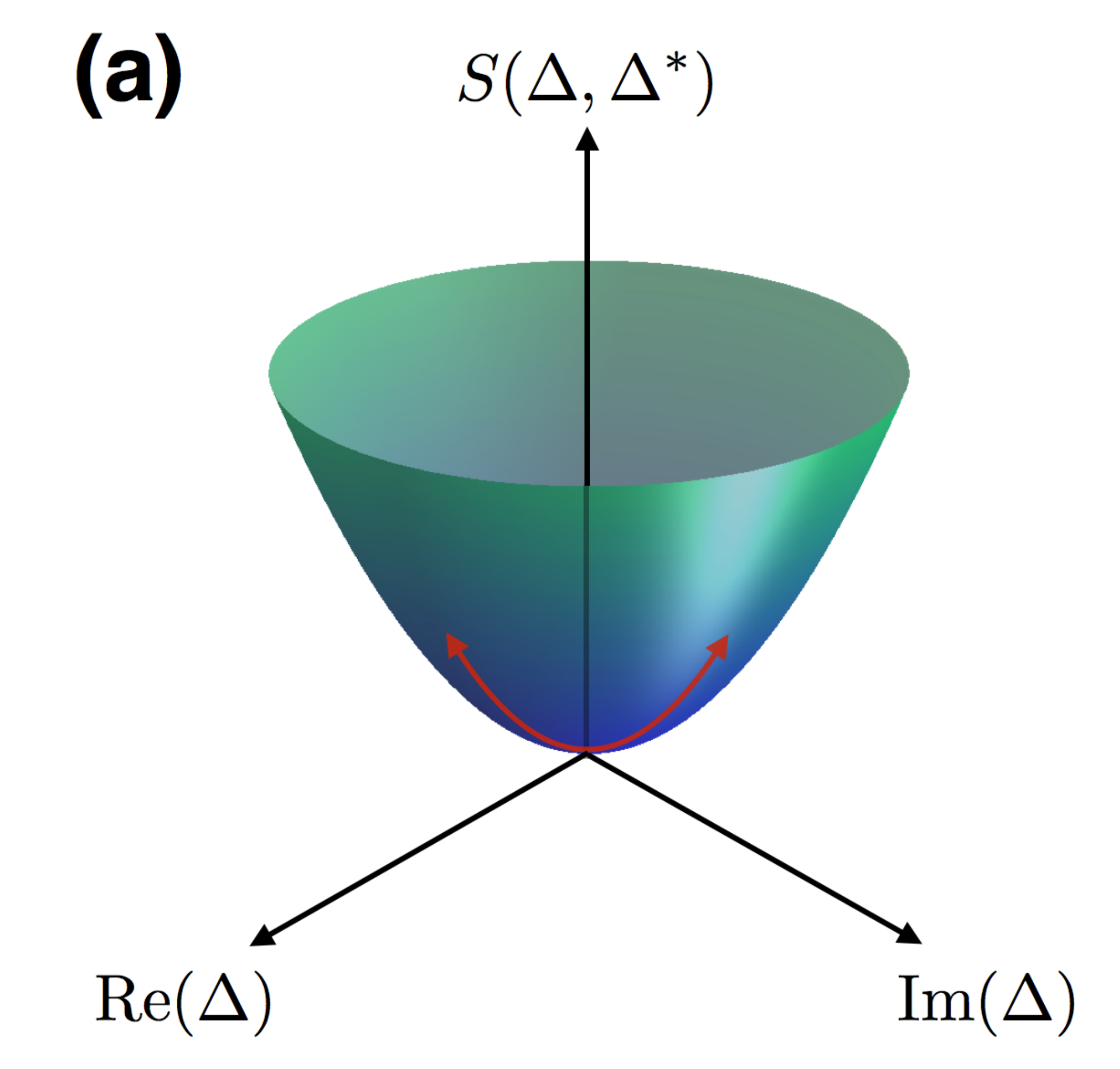}
\includegraphics[width=0.8\columnwidth]{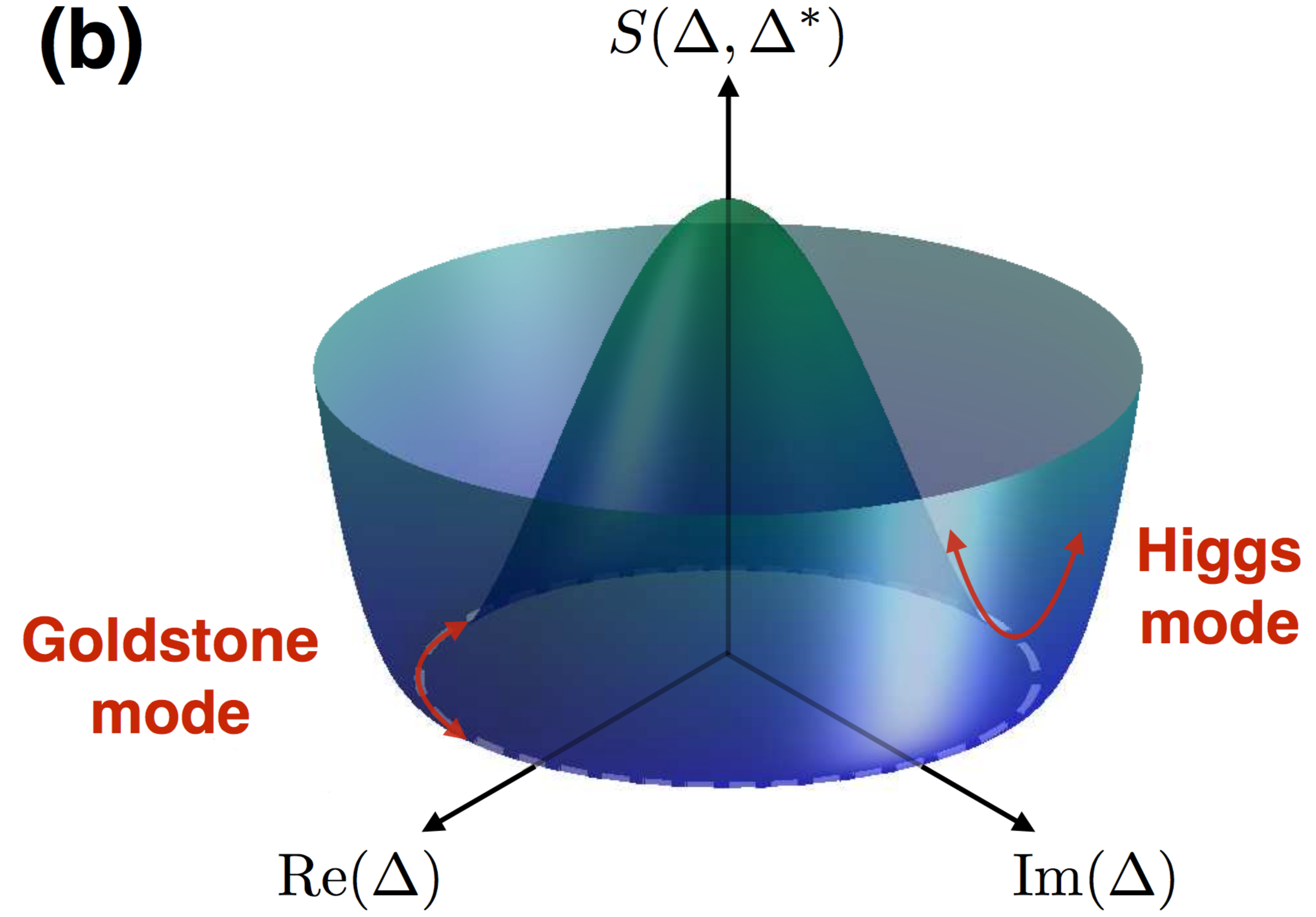}
\caption{ The free energy of the system as a function of the pairing field and the associated Higgs and Goldstone modes. (a) 
the weak coupling regime where the system is in the normal phase ($\epsilon_B< \epsilon_B^c$). (b) the strong coupling superfluid regime 
($\epsilon_B> \epsilon_B^c$). The Higgs  mode corresponds to amplitude
 fluctuations around the minimum of the effective action. 
At the quantum phase transition point between the two phases,  $\epsilon_B= \epsilon_B^c$, the Gaussian amplitude fluctuations around  $\Delta=0$ cost zero energy. 
The Goldstone mode corresponds to phase fluctuations in the superfluid phase. 
 } 
\label{MexicanFig}
\end{figure}
 
 Here, we develop a microscopic theory for the pairing of a  gas of fermionic atoms confined in a two-dimensional (2D) harmonic trap. The theory includes  
 the trapping potential exactly, and it is quantitatively reliable in the regime  where  the 
   pairing energy is smaller than the trap level spacing. We derive several analytical 
results for the pairing properties and the associated Higgs and Goldstone modes.
 In particular, we demonstrate that for certain  "magic numbers" of particles trapped, a sharp Higgs mode appears close to a quantum phase 
 transition between the normal and a superfluid phase. The trapping potential is shown to stabilise the Higgs mode  against  decay, since the 
 level spacing of the Goldstone spectrum is larger than the mode frequency close to the phase transition. 
  We demonstrate how the Higgs mode can be excited and investigated systematically in a new generation of experiments confining atoms in micro-traps,
   which are well suited to form 2D systems. 
  
\section{The system}
We consider a 2D system of fermionic atoms of mass $m$ with an equal number in two internal states denoted spin $\sigma=\uparrow,\downarrow$. The particles 
are  trapped in a circular symmetric potential $V(r)=m\omega_\perp^2r^2/2$ where $r^2=x^2+y^2$ and the temperature $T$ is zero.
The 2D confinement can be realised by a  tight trapping potential $m\omega_z^2z^2/2$  in the $z$-direction \cite{Martiyanov2010,Feld2011,Sommer2012,Zhang2012}
with $\omega_z$ larger than any other relevant energy. 
For low densities, only the short range $s$-wave interaction between particles with opposite spin  is important, and  it can  be modelled  by a contact interaction $g\delta({\mathbf r})$ 
with a high energy cut-off, where $\delta({\mathbf r})$ is the 2D delta function. 
 When the 3D scattering length $a$ is much smaller than  $l_z=1/\sqrt{m\omega_z}$, 
  $g$ can be obtained by integrating the 3D pseudopotential over the oscillator ground state in the $z$-direction yielding $g=\sqrt{2\pi}a/ml_z$. Here,
we will  eliminate $g$ in favor of the two-body binding energy to obtain results independent of the cut-off.

We focus on the  pairing properties of this  system using a functional approach, which highlights  the collective modes connected with 
pairing oscillations in a particularly  lucid way. 
Performing a Hubbard-Stratonovich transformation to introduce the pairing field $\Delta(x)$, the partition function of the system can be written as  
$\mathcal{Z}=\int {\mathcal D}(\Delta\Delta^*)\exp[-S(\Delta,\Delta^*)]$
 with  the action 
\begin{equation}
S(\Delta,\Delta^*)=-\rm{Tr}\log {\mathcal G}^{-1}-\frac 1 g \int_0^\beta d\tau \int d^2r|\Delta(x)|^2.
\label{Action}
\end{equation}
Here  $x=({\mathbf r},\tau)$ with $\tau$ the imaginary time, and  the inverse Green's function is 
\begin{equation}
{\mathcal G}^{-1}(x,x')=\begin{bmatrix}
\partial_\tau+H_0 & -\Delta(x)\\ 
-\Delta(x)^* & \partial_\tau-H_0
 \end{bmatrix}
\delta(x-x')
\label{Green}
\end{equation}
with $H_0=-\nabla^2/2m+V(r)-\epsilon_F$ the single particle Hamiltonian minus the Fermi energy $\epsilon_F$   (We take $\hbar=1$).
The trace in (\ref{Action}) is  over $x$ and spin space. We ignore the mean-field (Hartree) correction to the single particle energy since the density and therefore the 
Hartree potential is harmonic in the Thomas-Fermi approximation for a 2D gas. Consequently, it can  be included simply by renormalising 
the trapping frequency  $\omega_\perp$. Figure \ref{MexicanFig} illustrates the action $S(\Delta,\Delta^*)$ in the normal and superfluid phases, 
and  the corresponding Goldstone and Higgs modes. 

\section{Pairing and single particle properties} \label{Singleparticle}
The instability towards pairing is described using mean-field theory, which is obtained as usual from the  stationary phase approximation to ${\mathcal Z}$ yielding the 
Bogoliubov-de Gennes (BdG) equations~\cite{deGennes1989book}. We expand the  corresponding Bogoliubov wave-functions 
in the eigenfunctions $\phi_{nm}({\mathbf r})$ of the single particle Hamiltonian, i.e.\ 
 $H_0\phi_{nm}({\mathbf r})=\xi_n\phi_{nm}({\mathbf r})$ with $\xi_n=(n+1)\omega_\perp-\epsilon_F$,  
 $\phi_{nm}({\mathbf r})=R_{nm}(r)\exp(im\phi)/\sqrt{2\pi}$, 
 and $m=-n,-n+2,\ldots n$  the angular momentum along the $z$-axis. The ground state is circular symmetric corresponding to
 pairing between $\uparrow$ and $\downarrow$ atoms with opposite angular momentum. 
 The pairing between particles in shells $n$ and $n'$ is given by the matrix element  (see Appendix \ref{AppBdG})
 \begin{equation}
 \Delta_{nn'm}=\int_0^\infty drr R_{nm}(r)R_{n'm}(r)\Delta(r)
\label{Deltannm}
 \end{equation}
 with $\Delta({\mathbf r})=g\langle\psi_\downarrow({\mathbf r})\psi_\uparrow({\mathbf r})\rangle$ the mean-field pairing field. 
 Here, $\psi_\sigma({\mathbf r})$ is the field operator annihilating a particle with spin $\sigma$ at position ${\mathbf r}$. 
Since $\Delta(r)$ has a definite sign in the ground state whereas the sign of the radial functions $R_{nm}(r)$ in general oscillates, the intershell matrix 
 elements with $n'\neq n$ are suppressed compared to the intrashell matrix elements with $n'=n$ in (\ref{Deltannm}). 
We can therefore ignore the intershell matrix elements when the pairing energy is small compared to the trap level spacing $\omega_\perp$, i.e.\
 $\Delta_{nn'm}\ll \omega_\perp$. We refer to this regime as \emph{intrashell pairing}, since the Cooper pairs are formed within each harmonic oscillator shell. 
The BdG equations then simplify into  $2\times 2$ matrix equations for each pair of quantum numbers $(n,m)$, with the  solutions  
$E_{nm}=(\xi_n^2+\Delta_{nnm}^2)^{1/2}$, $u_{nm}^2=(1+\xi_n/E_{nm})/2$, and $v_{nm}^2=(1-\xi_n/E_{nm})/2$. In addition, 
 the $m$-dependence of $\Delta_{nnm}$ is weak which can be shown explicitly using the Thomas-Fermi approximation (see Appendix \ref{AppBdG}). 
 We therefore make the approximation 
 $\Delta_{nnm}\simeq \Delta_n\equiv\sum_{m=-n}^n\Delta_{nnm}/\Omega_n$, where $\Omega_n=n+1$ is the degeneracy of the $n$'th shell. 
Using that the main contribution to the pairing is from the shells around  the Fermi level, we end up with the gap equation (see Appendix \ref{AppBdG})
 \begin{equation}
-\frac 1 G=\sum_{n}\frac{1}{2E_{n}}=\sum_{n}\frac{1}{2\epsilon_n-\epsilon_{\rm tb}}
\label{GapEqn2}
\end{equation}
where  $E_n=(\xi_n^2+\Delta_n^2)^{1/2}$ with $\Delta_n=\Delta/\sqrt{\Omega_n}$, and $\epsilon_n=(n+1)\omega_\perp$. The effective coupling strength is 
\begin{equation}
G=2\pi g \frac{\int_0^\infty dr  r\rho_{n_F}(r)^2}{\Omega_{n_F}},
\label{Gdef}      
\end{equation}
where $n_F$ denotes the highest occupied shell in the normal phase, and 
$\rho_n(r)=\sum_mR_{nm}(r)^2/2\pi $ is the particle density of a completely filled shell $n$.  

In the second equality in  (\ref{GapEqn2}), we have used that the energy $\epsilon_{\rm tb}$ of a 
two-body  state is determined by 
 $1/G=-\sum_{n}1/({2\epsilon_n-\epsilon_{\rm tb}})$ as explained in  Appendix \ref{AppTB}.
 In the perturbative regime, this yields  $\epsilon_{\rm tb}=2\epsilon_n+G=2\epsilon_n+\sqrt2a\omega_\perp/\sqrt\pi l_z$,
 where we have used the Thomas-Fermi result $G=g/2\pi l^2$  and $g=\sqrt{2\pi}a/ml_z$ for weak confinement $a\ll l_z$ (see Appendix \ref{AppBdG}). 
 This recovers the exact result for the two-body binding energy of an $s$-wave state in a 
 2D Harmonic trap  in the perturbative regime~\cite{Busch1998}, which illustrates an important point: The approximations 
 we make are systematic, and  (\ref{GapEqn2}) is it not merely a schematic model -- it
 provides an accurate description of the correlations with monopole symmetry in the instrashell regime. 
 By replacing $1/G$ by the two-body energy, we have arrived at quantitative reliable theory for the monopole pairing correlations,  
 which is well-defined for an infinite cut-off. 
  It represents a crucial simplification   which allows us to derive several analytical results.  For a 3D spherical trap, 
  similar approximations were shown to yield very accurate results when compared to a full solution of the BdG equations~\cite{Bruun2002a}.

There are two qualitatively different cases for pairing: The open shell case $\epsilon_F=(n_F+1)\omega_\perp$
 where the highest occupied shell $n=n_F$ is partly filled, 
 and the "magic number" case  $\epsilon_F=(n_F+3/2)\omega_\perp$ with a completely filled highest shell $n=n_F$. 
 We parametrise the interaction strength by the  two-body binding energy $\epsilon_{B}>0$ per particle, defined as   $\epsilon_{\rm tb}=2\omega-2\epsilon_{B}$. 
 To obtain analytical results, we   expand  (\ref{GapEqn2}) in $\Delta/\omega_\perp$ and 
$\epsilon_B/\omega_\perp$. For the open shell case, this yields after evaluating the sums 
\begin{equation}
 \Delta_{n_F}=\frac{\epsilon_B}{1-\epsilon_B(\gamma+\ln n_F)/\omega_\perp},
 \label{OpenshellGap}
 \end{equation}
where $\gamma=0.577$ is the Euler-Mascheroni constant and we have ignored $1/n_F$ corrections. 

\begin{figure}
\includegraphics[width=0.99\columnwidth]{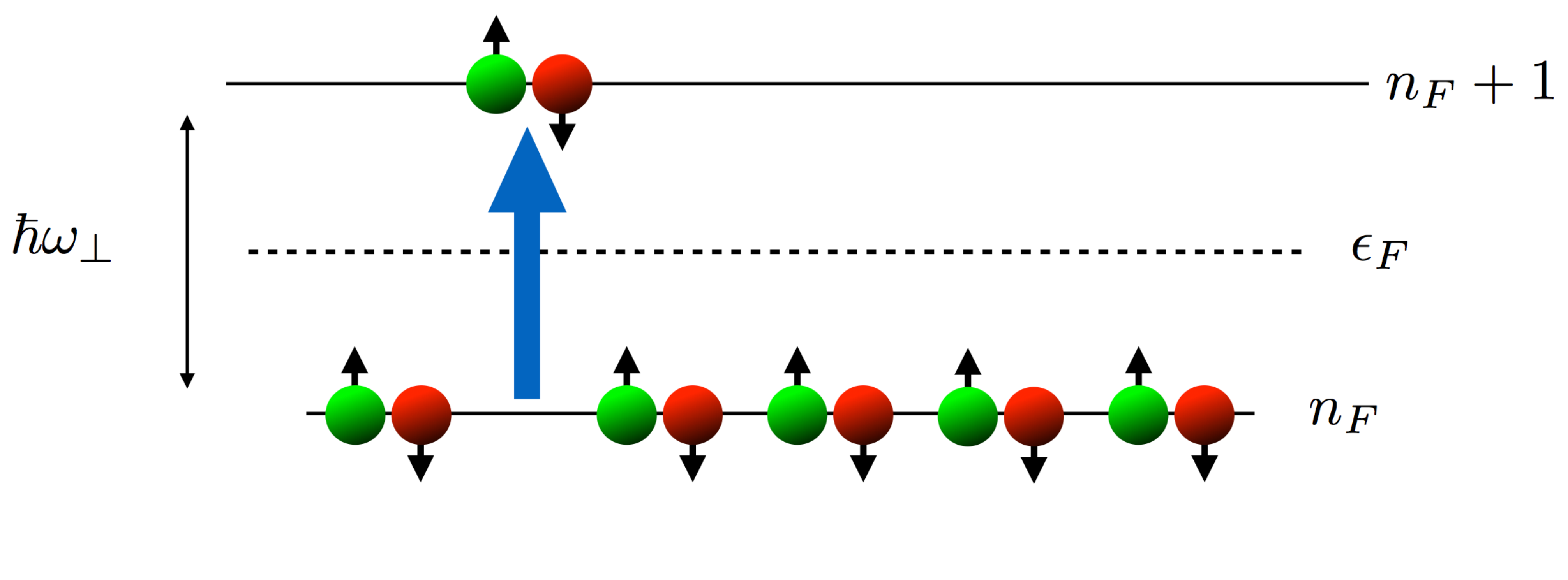}
\caption{For the closed shell case, there is only pairing for strong enough attraction, $\epsilon_B>\epsilon_B^c$, so that is is energetically favourable to excite 
Cooper pairs from the highest occupied to the lowest empty shell. The collective mode in the normal phase  is formed by making   coherent 
 excitations pairs across  the chemical, and the mode frequency  goes to zero when the system becomes unstable to pairing.  
}
\label{PairExciteFig}
\end{figure}

For the closed shell case $\epsilon_F=(n_F+3/2)\omega_\perp$, there is only pairing for strong enough attraction when it is energetically
 favourable to excite pairs from the highest filled shell  $n=n_F$ to the lowest empty shell $n=n_F+1$ as illustrated in Fig.\ \ref{PairExciteFig}.
 Expanding  (\ref{GapEqn2})  in $\Delta/\omega_\perp$ and 
$\epsilon_B/\omega_\perp$ yields 
 \begin{equation}
 \frac{\epsilon^c_B}{\omega_\perp}=\frac{B(n_F)}{2\xi(2)}[\sqrt{1+4\xi(2)/B( n_F)^2}-1]
 \label{ClosedshellCritical}
 \end{equation}
for the critical attraction strength for pairing  with $B(n_F)=\gamma+4\ln 2+\ln n_F$ and $\xi(z)$ Riemann's zeta function. 
For $\epsilon_B>\epsilon_B^c$, we obtain 
\begin{equation}
 \Delta_{n_F}=\frac{\omega_\perp}{\sqrt{7\xi(3)}}\sqrt{\frac{\omega_\perp}{\epsilon^c_B}-\frac{\omega_\perp}{\epsilon_B}+
 \xi(2)\left(\frac{\epsilon_B}{\omega_\perp}-\frac{\epsilon^c_B}{\omega_\perp}\right)}
 \label{ClosedshellGap}
 \end{equation}
for the pairing energy. 

\section{Collective modes}
We now analyse the collective modes arising from the fluctuations $\delta\Delta(x)$ of the pairing field around the  mean-field solution. 
These fluctuations are included by writing $\mathcal{G}^{-1}=\mathcal{G}_{\rm mf}^{-1}-\Sigma$ where $\mathcal{G}^{-1}_{\rm mf}$
 is given by (\ref{Green}) using the mean-field pairing field,
and  
\begin{equation}
\Sigma(x)=\begin{bmatrix}
0 & \delta\Delta(x)\\ 
\delta\Delta(x)^* & 0
 \end{bmatrix}
\end{equation}
contains the fluctuations. We have   
$\rm{Tr}\log \mathcal{G}^{-1}=\rm{Tr}\log \mathcal{G}_{\rm mf}^{-1}-\sum_{n=1}^\infty\rm{Tr}[(\mathcal{G}_{\rm mf}\Sigma)^n]/n$, and 
since the linear term vanishes as we expand around a stationary point,  we obtain   
$S(\Delta,\Delta^*)\simeq -\rm{Tr}\log \mathcal{G}_{\rm mf}^{-1}+S_2(\delta\Delta,\delta\Delta^*)$
to quadratic order. Here  $S_2(\delta\Delta,\delta\Delta^*)={\rm Tr[\mathcal{G}_{\rm mf}\Sigma \mathcal{G}_{\rm mf}\Sigma]}/2-\int d^3r|\delta\Delta(x)|^2/g$.
As explained in Appendix \ref{AppColl},  evaluating the trace and the resulting Matsubara sums yields  
 $S_2=T\sum_l[s_2(i\omega_l)-\int d^2r\mathbf{d}({\mathbf r},i\omega_l)^\dagger\mathbf{d}({\mathbf r},i\omega_l)/g]/2$ with 
  $\mathbf{d}({\mathbf r},i\omega_l)^\dagger=[\delta\Delta({\mathbf r},i\omega_l)^*,\delta\Delta({\mathbf r},-i\omega_l)]$ and $\omega_l=2l\pi T$  
  with $l=0,\pm 1, \pm 2\ldots$ a Bose Matsubara frequency.

To find the collective modes, we analytically continue to real time $i\omega_l\rightarrow\omega+i0_+$.
 The collective mode frequencies $\omega$ are then obtained by finding the  zeroes in the inverse pair fluctuation propagator $s_2(\omega)$.
For low energy, the collective modes  split into phase and amplitude fluctuations, and 
 we therefore write $\delta\Delta({\mathbf r},t)=[\kappa({\mathbf r},t)+i\theta({\mathbf r},t)]/\sqrt 2$ where $t$ is  time~\cite{Engelbrecht1997}. 
Both $\kappa({\mathbf r},t)$ and $\theta({\mathbf r},t)$ are real and they  describe amplitude (Higgs) and phase (Goldstone) fluctuations respectively. 
  We find 
\begin{equation}
s_2(\omega)=\int d^2rd^2r'\mathbf{v}({\mathbf r},\omega)^\dagger
{\mathbf A}({\mathbf r},{\mathbf r}',\omega)
\mathbf{v}({\mathbf r}',\omega)
\label{s2}
\end{equation} 
where ${\mathbf A}({\mathbf r},{\mathbf r}',\omega)=\sum_{nm}|\phi_{nm}({\mathbf r})|^2{\mathbf A}(n,\omega)|\phi_{nm}^*({\mathbf r}')|^2$,  
\begin{gather}
A_{11}(n,\omega)=\frac{\xi_n^2}{2E_n(\omega^2-4E_n^2)}\hspace{0.5cm}A_{22}(n,\omega)=\frac{2E_n}{\omega^2-4E_n^2}\nonumber\\
A_{12}(n,\omega)=-A_{21}(n,\omega)=i\omega\frac{u_n^4-v_n^4}{4E_n^2-\omega^2},
\label{Amatrixelements}
 \end{gather}
 and ${\mathbf v}({\mathbf r},\omega)^\dagger=[\kappa({\mathbf r},\omega)^*,\theta({\mathbf r},\omega)^*]$.

\subsection{The Goldstone mode}
The Goldstone mode is found by solving $\int d^2r'{\mathbf A}_{22}({\mathbf r},{\mathbf r}',\omega)\theta({\mathbf r}')=\theta({\mathbf r})/g$.
We take  a circular symmetric eigenfunction $\theta(r)$ corresponding to a monopole mode. In the spirit of the  intra-shell regime, we ignore the weak $n,m$-dependence of the resulting integrals, 
 when ${\mathbf A}({\mathbf r},{\mathbf r}',\omega)$ is expressed in terms of the functions $\phi_{nm}({\mathbf r})$. 
  As detailed in Appendix \ref{AppColl}, this gives the eigenvalue equation 
 \begin{equation}
\sum_n\frac{2E_n}{4E_n^2-\omega^2}=\sum_{n}\frac{1}{2\epsilon_n-\epsilon_{\rm tb}}
 \label{GoldstoneMode}
 \end{equation} 
 for the collective Goldstone mode, where we again have used  the two-body energy to eliminate the  coupling constant $G$.
 Using the gap equation (\ref{GapEqn2}), we see that $\omega=0$ is a solution to (\ref{GoldstoneMode}). 
Since $A_{12}({\mathbf r},{\mathbf r}',\omega)\propto \omega$ and  $A_{21}({\mathbf r},{\mathbf r}',\omega)\propto\omega$, 
this solution is furthermore completely decoupled from the amplitude oscillations. 
 The theory thus recovers  the zero energy Goldstone mode corresponding to the broken $U(1)$  symmetry of the phase of the pairing field. 

\subsection{The Higgs mode}
The Higgs amplitude mode is found by solving $\int d^2r'{\mathbf A}_{11}({\mathbf r},{\mathbf r}',\omega)\kappa({\mathbf r}')=\kappa({\mathbf r})/g$.
Taking $\kappa({\mathbf r})$ to be circular symmetric and ignoring again the weak $n,m$ dependence of the integrals yields 
an eigenfuction of the form $\kappa({\mathbf r})\propto\sum_nA_{11}(n,\omega)\rho_n({\mathbf r})$. We obtain (see Appendix \ref{AppColl})
\begin{equation}
 \sum_n\frac{2\xi_n^2}{E_n(4E_n^2-\omega^2)}=\sum_{n}\frac{1}{2\epsilon_n-\epsilon_{\rm tb}}.
 \label{HiggsMode}
 \end{equation} 
Assuming perfect particle-hole symmetry around the Fermi level,  we see from the gap equation (\ref{GapEqn2})
 that  $\omega=2\Delta_{n_F}$ is a solution. This mode is decoupled from the 
 Goldstone modes and therefore undamped, since it  follows from  
 (\ref{Amatrixelements}) that $\sum_n{\mathbf A}_{12}(n,2\Delta_{n_F})=\sum_n{\mathbf A}_{21}(n,2\Delta_{n_F})=0$ for 
 perfect particle-hole symmetry.  
 This is the undamped  Higgs mode 
for the superfluid phase  corresponding to monopole amplitude oscillations  of the pairing field. Note that  particle-hole symmetry 
is  equivalent to a Lorentz-invariant low energy effective theory~\cite{Varma2002}.

Including the particle-hole asymmetry around the Fermi level  changes the mode frequency only by a small amount when  $n_F$ is not too small, as will be confirmed 
numerically below. In addition, it leads to damping of the Higgs mode due to coupling to the Goldstone modes. 
This damping is however weak for two reasons. First, in the
 intrashell regime pairing mainly occurs in the shells around the Fermi energy 
 which are approximately particle-hole symmetric leading to a weak coupling to the Goldstone modes.   Second, the Goldstone modes correspond to 
 density oscillations with a typical energy $\sim \omega_\perp$ both in the collisionless and in the hydrodynamic 
 regimes~\cite{Pitaevskii1997,Gao2012,Baur2013}.  The  Higgs mode is therefore well separated in energy from these modes since $\Delta\ll \omega_\perp$ 
 in the intrashell regime. 
In the open shell case, there are however low lying pair breaking excitations with energy $\sim2\Delta$ which can damp 
 the Higgs mode, but in the closed shell case, no such low energy modes exist since the lowest single particle energies are 
 $(\omega_\perp^2/4+\Delta_{n_F}^2)^{1/2}$.  The existence of a  well defined
  Higgs mode with a sharp spectral peak for the closed shell case is one of the main 
  results of this paper. The lack of damping of this mode   is a major advantage compared to previous experimental table-top realisations 
  of the Higgs mode, where  coupling to a continuum  of  modes leads to significant damping~\cite{Podolsky2011,Endres2012}.  
 
 \subsection{Normal phase}
In the closed shell case,  the system is in the normal phase for weak attraction 
with $\epsilon_B<\epsilon_B^c$. In this case, (\ref{HiggsMode})  has to be solved with 
$E_n=|\xi_n|$ and the   amplitude modes correspond to coherently either 
 adding a pair of particles,  or  removing a pair of particles.
 The particle conserving collective modes correspond to subsequently adding and removing a pair of particles, and their 
 frequencies is therefore \emph{twice} the frequency  obtained by solving (\ref{HiggsMode}). 
In the  weak coupling regime $\epsilon_B\ll \epsilon_B^c$, we obtain the collective mode frequency 
 $\omega=2\omega_\perp-4\epsilon_B$. This result can also be derived directly from first order perturbation 
 theory using $|E\rangle=\Gamma_{n_F+1}^\dagger\Gamma_{n_F}|G\rangle$ for the excited state, where 
 $\Gamma_n^\dagger=\sum_{m}a_{nm\uparrow}^\dagger a_{n-m\downarrow}^\dagger/\sqrt{\Omega_n}$ and $|G\rangle$ is the non-interacting ground state 
 with all shells up to and including $n_F$ completely filled.  The state $|E\rangle$ is formed by   exciting $\uparrow\downarrow$ pairs of 
  from the highest fully occupied shell $n_F$ to the lowest unoccupied 
shell $n_F+1$. The excitation energy initially decreases  with increasing attraction
 since the particles can increase their overlap in the excited state. When the excitation energy 
goes to zero, the system can spontaneously excite pairs from the shell $n_F$ to the shell $n_F+1$, and it is unstable towards 
Cooper pair formation, see Fig.\ \ref{PairExciteFig}.
   Close to critical coupling strength for pairing  $\epsilon_B\lesssim\epsilon_B^c$,
we expand (\ref{HiggsMode}) in $\epsilon_B/\omega_\perp$ and $\omega/\omega_\perp$. Evaluating the resulting sums to accuracy 
$1/n_F$ yields
\begin{equation}
 \frac{\omega}{\omega_\perp}=\frac 2{\sqrt{7\xi(3)}}\sqrt{\frac{\omega_\perp}{\epsilon_B}-\frac{\omega_\perp}{\epsilon^c_B}+
 \xi(2)\left(\frac{\epsilon^c_B}{\omega_\perp}-\frac{\epsilon_B}{\omega_\perp}\right)}.
 \label{HiggsModeNormalClose}
 \end{equation}
 
 \section{Numerical results}
\begin{figure}
\includegraphics[width=0.9\columnwidth]{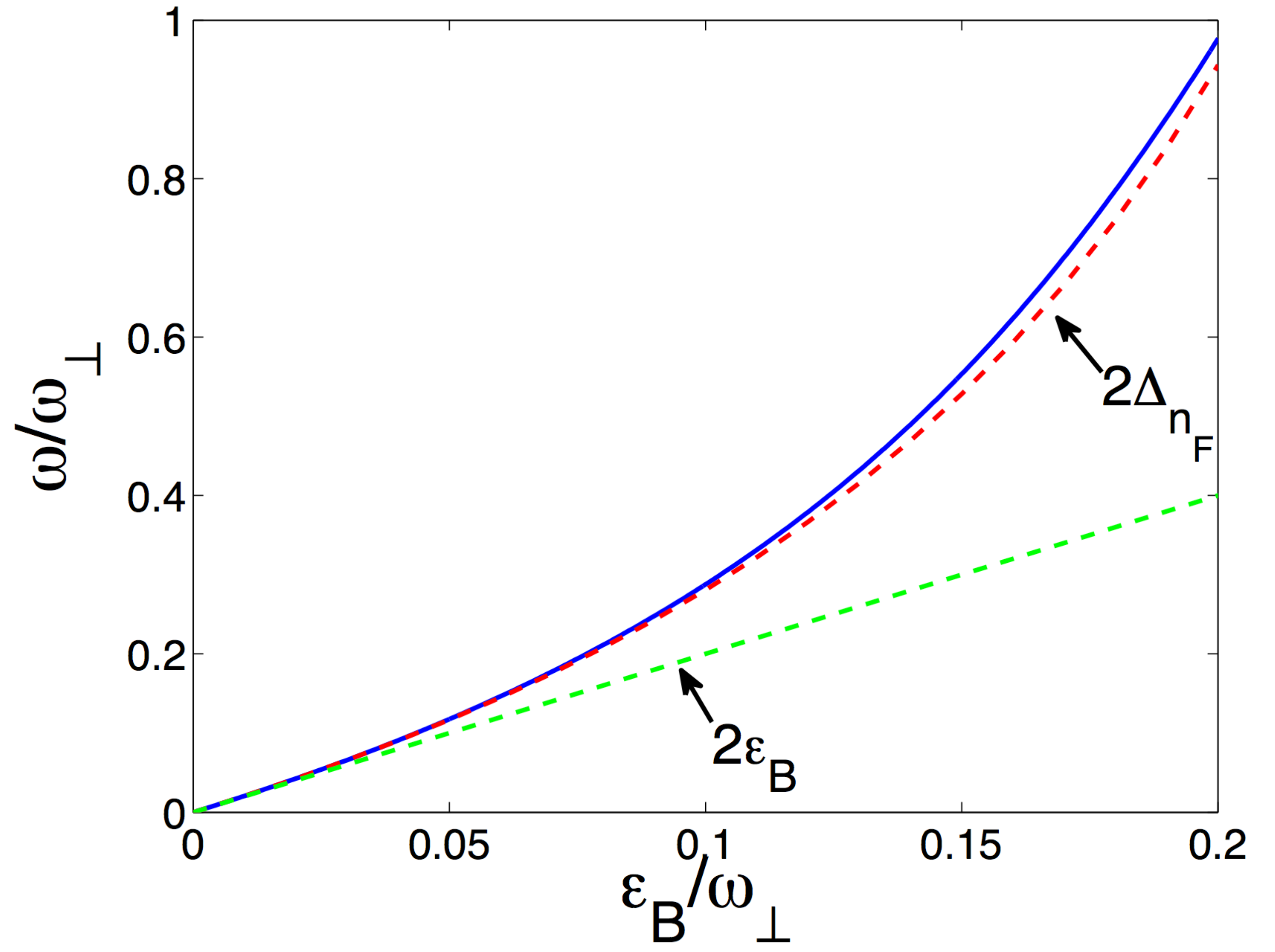}
\caption{The Higgs mode for the open shell case. We plot the Higgs amplitude mode energy
  as a function of the  two-body binding energy per particle $\epsilon_B$
for the open shell case with $n_F=10$. The  solid line is obtained by numerically solving (\ref{HiggsMode}),  the dashed   line is $\omega=2\Delta_{n_F}$
with $\Delta_{n_F}$ given by  (\ref{OpenshellGap}), 
and the dotted line is $\omega=2\epsilon_B$.   
} 
\label{OpenShellFig}
\end{figure}

The Higgs  mode frequency obtained from (\ref{HiggsMode}) with $n_F=10$ is shown in Fig.\ \ref{OpenShellFig} for the open shell case. 
We see that there is very good agreement between the numerical solution and the 
analytical results. The numerical solution of (\ref{HiggsMode}) is essentially indistinguishable from $2\Delta_{n_F}$ with $\Delta_{n_F}$ determined from 
(\ref{OpenshellGap}). This agreement shows that  particle-hole asymmetry  has a negligible effect on the collective mode frequency
as well as on the single particle pairing, so that that (\ref{OpenshellGap}) is an  accurate expression for the solution to the gap equation (\ref{GapEqn2}). 
 The pairing and therefore the Higgs mode energy  grows linearly with the binding energy for weak coupling whereas it 
increases more quickly when more shells participate in the pairing. 
In the case of smaller particle numbers, i.e.\ smaller $n_F$, the analytical results agrees less with the numerics, since particle-hole asymmetry 
($1/n_F$ effects) become larger.  

\begin{figure}
\includegraphics[width=0.9\columnwidth]{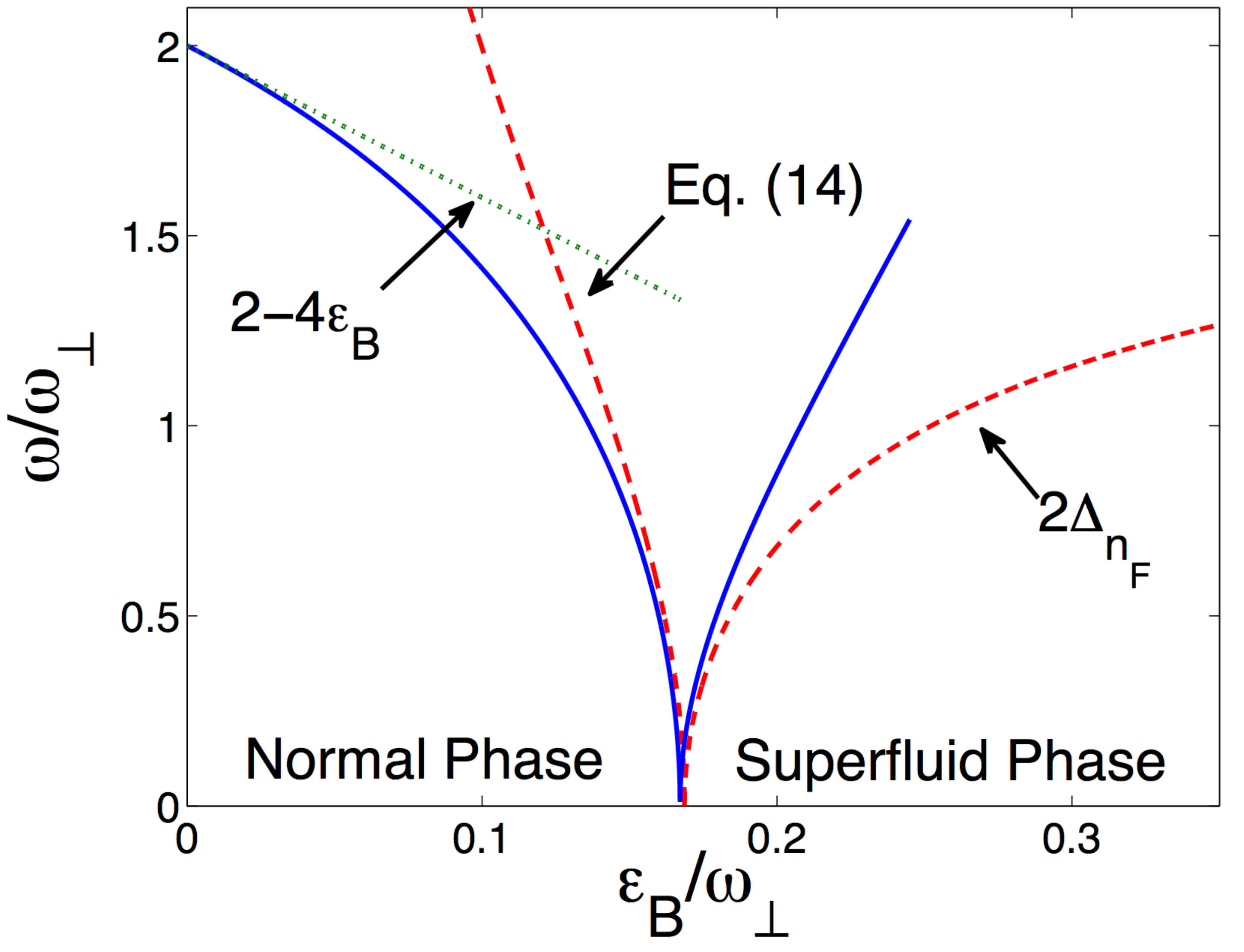}
\caption{The Higgs mode  for the closed shell case. We plot the Higgs amplitude mode energy  
as a function of the  two-body binding energy per particle $\epsilon_B$
for the closed  shell case with $n_F=10$. The  solid line is obtained by numerically solving (\ref{HiggsMode}), and   the dashed   lines are (\ref{HiggsModeNormalClose})
 for   $\epsilon_B<\epsilon_B^c$, and $\omega=2\Delta_{n_F}$
with $\Delta_{n_F}$given by  (\ref{ClosedshellGap})
for $\epsilon_B>\epsilon_B^c$. The critical two-body binding energy 
 $\epsilon_B^c$ is given by (\ref{ClosedshellCritical}). The dotted line is the perturbative 
result $\omega=2\omega_\perp-4\epsilon_B$. } 
\label{ClosedShellFig}
\end{figure}

In Fig.\ \ref{ClosedShellFig}, we plot the Higgs mode frequency  for the closed shell case  with $n_F=10$. 
Again, we see that the analytical formulas agree very well with the numerical solution
showing that particle-hole asymmetry has a negligible effect on pairing  and the Higgs mode frequency in the intrashell regime. As for the open shell case, 
the agreement is less for smaller $n_F$.   
The energy initially decreases from non-interacting value $2\omega_\perp$ with increasing binding $\epsilon_B$.  At the 
critical coupling strength for pairing, the mode has zero frequency and for stronger attraction when the system is superfluid, the 
frequency is $2\Delta_{n_F}$.  This characteristic non-monotonic behaviour of the Higgs frequency near the quantum phase transition  has a clear 
interpretation in terms of the effective interaction and broken symmetry (see Fig.\ \ref{MexicanFig}), and it provides a smoking 
gun signal for the Higgs mode. 

We note that  the intrashell ansatz eventually breaks down when  $\Delta_{n_F}$ is comparable to $\omega_\perp$ and matrix elements with $n'\neq n$
in (\ref{Deltannm}) become important. This corresponds to the coherence length of the Cooper pairs becoming smaller than the system 
size. The  system then approaches the bulk limit where the Higgs mode becomes damped by coupling to the Goldstone modes.  In the case of a 3D harmonic trap, a comparison with a full solution to the BdG equations shows that 
the breakdown of the intrashell regime occurs for $\Delta_N\simeq \omega_\perp/2$ both at the single particle level~\cite{Bruun2002a}, 
and for the collective modes~\cite{Bruun2002b,Bruun2001}.

 \section{Experimental realisation}
The Higgs amplitude mode does not couple strongly to density oscillations  which makes it hard to observe in condensed matter systems~\cite{Varma2002}. 
For cold atoms it can on the other hand be excited rather straightforwardly by modulating the coupling strength between the atoms for instance by changing the external magnetic field close 
 to a Feshbach resonance. This leads to a time-dependent interaction  of the form  $H_{\rm int}=G(t)\sum_{nn'}\Gamma_n^\dagger \Gamma_{n'}$,
 which couples strongly to the Higgs mode by exciting pairs across the Fermi energy, see Fig.\ \ref{PairExciteFig} and Appendix \ref{AppTB}.
  When $G(t)$ is modulated at the resonance frequency, the Higgs mode will be excited which can be detected for instance  by 
 measuring the energy transferred to the system. For a small system, the transferred energy  should be sizable fraction of the total energy. 
 Alternatively,  the Higgs mode 
 could be detected by counting the number of atoms excited to  higher shells. The number of atoms in individual oscillator levels was recently 
  counted with single atom precision using a new generation of  micro traps~\cite{Serwane2011,Zurn2012}. 
  The trapping frequencies are  of the order  $\sim 10$kHz in these  experiments, which  
  means that for  $T\ll 500$nK, effects of a non-zero temperature are small and one should be able to observe the effects described in this paper. 
  Finally, while present micro traps realise 1D systems with less than 10 particles, it is experimentally feasible  to make these systems 2D,
   which in addition could increase the particle number~\cite{WeitenbergJochim2014}. 
   
   A small particle number will give rise to significant finite size effects such as 
   the lack of a sharp quantum phase transition where the Higgs mode frequency goes to zero. Instead, the finite size version of the Higgs mode will be 
   characterised by   a smooth non-monotonic frequency as a function of the attraction: First, it will  
   decrease  with increasing attraction until it reaches a differentiable minimum  of non-zero frequency, after which it  increases with increasing attraction. The minimum frequency will  
   decrease with increasing particle number becoming sharper and sharper  at the same time, as the system approaches the many-body limit.
   This transition between few- and many-body dynamics  will be very interesting to observe.

 \section{Conclusions and outlook}
We developed a microscopic theory for pairing and collective modes in a trapped 2D Fermi gas, which takes 
the trapping potential into account exactly. The theory is  quantitatively reliable when the pairing energy is much smaller than the trap level spacing, and at  the same it 
is simple enough to allow the derivation of  several analytical results. Using this theory, we demonstrated the existence of a sharp  
Higgs mode close to the  quantum phase transition between a normal and a superfluid phase, when the system is in a closed 
shell configuration. The trapping confinement was shown to  stabilise the Higgs mode against decay, 
since it makes the Goldstone spectrum discrete with a level spacing much larger than the Higgs energy. We then discussed  how 
a new generation of cold atom experiments using micro traps can realise the physics described in this paper. 
 
 Our results open up the intriguing prospect of using cold atoms in micro traps to observe for the first time a long lived Higgs mode in a confined geometry. 
 This is relevant to certain models in high energy physics~\cite{Randall1999,Arkani-Hamed1998}, 
 as well as to the so-called pair vibration modes, which play a central role in the theory of 
atomic nuclei~\cite{BohrMottelson1975book,Potel2013}. In atomic nuclei, it is however very challenging  to 
calculate the properties of these modes microscopically, and  they are furthermore probed rather indirectly in  nucleon transfer 
reactions whose interpretation is  subject  to intense debate. Finally, we discussed how  micro traps can be used to investigate the fundamental
 question how the "standard model" of  broken symmetry and collective modes in condensed matter physics 
emerges, as the dynamics changes from few- to many-body physics with increasing particle number~\cite{Sachdev2011book,Wenz2013}. 

\acknowledgements
It is a pleasure to thank C.\ Weitenberg, S.\ Jochim, P.\ Massignan, H.\ Fynbo, and K.\ Riisager  for  useful discussions. Financial support from the
 Villum Foundation via grant VKR023163, and the ESF POLATOM network is acknowledged. 
\appendix
\section{The Bogoliubov-de Gennes equations}\label{AppBdG}
We expand the Bogoliubov wave-functions $[u_m^\eta({\mathbf r}),v_m^\eta({\mathbf r})]$
in the eigenfunctions  of the single particle Hamiltonian,  i.e. $u^\eta_m({\mathbf r})=\sum_nu^\eta_{nm}\phi_{nm}({\mathbf r})$ and likewise for $v^\eta_m({\mathbf r})$.
In this basis, the Bogoliubov-de Gennes equations read 
\begin{equation}
E^\eta_{m}
\begin{bmatrix}
u^\eta_{nm}\\
v^\eta_{nm}
\end{bmatrix}
=\sum_{n'}
\begin{bmatrix}
\xi_{n}\delta_{n,n'} & \Delta_{nn'm}\\ 
\  \Delta_{nn'm} & -\xi_{n}\delta_{n,n'}
 \end{bmatrix}
\begin{bmatrix}
u^\eta_{n'm}\\
v^\eta_{n'm}
\end{bmatrix}
\label{BdG}
\end{equation}
for the quasiparticle energies $E_m^\eta$. The pairing matrix element is given by (\ref{Deltannm}). Since the $m$-dependence is weak, we make 
the approximation 
 $\Delta_{nnm}\simeq \Delta_n\equiv\sum_{m=-n}^n\Delta_{nnm}/\Omega_n$, and the self-consistent gap equation (\ref{Deltannm})  becomes 
 \begin{equation}
\Delta_n=-g \sum_{n'}\frac{\Delta_{n'}}{2E_{n'}}\frac{\int_0^\infty dr r \rho_n(r)\rho_{n'}(r)}{\int_0^\infty drr \rho_n(r)}
\label{GapEqn1}
\end{equation}
where  we have used 
$\Omega_n=\sum_m\int_0^\infty dr rR_{nm}(r)^2$. From (\ref{GapEqn1}), we can define an appropriately symmetrized effective coupling strength 
between the shells $n$ and $n'$ as  
\begin{equation}
G_{nn'}=2\pi g \frac{\int_0^\infty dr  r\rho_{n}(r)\rho_{n'}(r)}{\sqrt{\Omega_n\Omega_{n'}}}.
\end{equation}
The effective coupling strength depends  only weakly on $n$ and $n'$ for the shells around the Fermi energy  
which contribute most to the pairing.  We therefore write $G_{nn'}\simeq G_{n_Fn_F}\equiv G$, and the gap equation  (\ref{GapEqn1})  simplifies to
(\ref{GapEqn2}).
The pairing field  corresponding to this solution is 
\begin{equation}
\Delta(r)=-g  \sum_{n}\frac{\Delta_{n}}{2E_{n}}\rho_n(r).
\label{Deltar}
\end{equation}

The weak $m$-dependence of the pairing can be checked by invoking the Thomas-Fermi approximation. 
 In the Thomas-Fermi regime  $n\gg 1$, we have $\rho_{n}(r)=\partial_n\rho(r)=1/2\pi l^2$ 
for $r< \sqrt{2n+1}l$ and $\rho_{n}(r)=0$ for $r> \sqrt{2n+1}l$ with $l=1/\sqrt{m\omega_\perp}$.
 Here,  $\partial_{n}\rho(r)$ is the derivative of the total (single spin) density $\rho(r)$ of filled shells up to and including 
$n$, with respect to $n$. We have $\rho(r)=(n_F+1)(1-r^2/R_{TF}^2)/2\pi l^2$ for a 2D gas at $T=0$ with $T_{TF}=\sqrt{2n_F+1}l$. Using this in (\ref{Gdef}), we obtain 
 $G=g/2\pi l^2$ in the Thomas-Fermi regime. For weak coupling, when only the highest occupied
 shell $n=n_F$ contributes to the pairing we then obtain from (\ref{Deltar}) $\Delta(r)=-g/4\pi l^2$ for $r<R_{TF}$, i.e. 
   $\Delta(r)$ is simply a constant. It 
 follows that  $\Delta_{nnm}=\int_0^\infty drr R_{nm}(r)^2\Delta(r)$ is indeed independent of  $m$ making our assumption self-consistent.
 In general, the highest occupied shell dominates pairing in the intrashell regime which explains why 
 the pairing depends only weakly on  $m$. 

\section{Two-body energy}\label{AppTB}
The effective Hamiltonian describing the monopole correlations in the intrashell regime is 
\begin{equation}
 H_{\rm eff}=\sum_{nm\sigma}\epsilon_n a_{nm\sigma}^\dagger a_{nm\sigma}+G\sum_{nn'}\Gamma_n^\dagger \Gamma_{n'}
\end{equation}
where  $\Gamma_n^\dagger=\sum_{m}a_{nm\uparrow}^\dagger a_{n-m\downarrow}^\dagger/\sqrt{\Omega_n}$ and $a_{nm\sigma}$ removes
 a particle in state $(n,m)$ with spin $\sigma$. It is easy to show that this Hamiltonian leads to the gap equation (\ref{GapEqn2}). 
Writing the two-body state as $\sum_nc_n\Gamma_n^\dagger|0\rangle$ with $|0\rangle$ the vacuum state, 
it follows that the two-body energy is given by 
 \begin{equation}
\frac 1 G=-\sum_{n}\frac 1 {2\epsilon_n-\epsilon_{\rm tb}}.
  \label{twobody}
 \end{equation}

\section{Gaussian fluctuations}\label{AppColl}
We evaluate the trace ${\rm Tr}[\mathcal{G}_{\rm mf}\Sigma \mathcal{G}_{\rm mf}\Sigma]$ by going to Matsubara space. 
The $ij$'th component of the Green's function is in the intrashell regime 
given by $\mathcal{G}_{{\rm mf},ij}(\mathbf{r},\mathbf{r}',i\omega_j)=\sum_{nm}\phi_{nm}(\mathbf{r}')\mathcal{G}_{{\rm mf},ij}(n,i\omega_j)\phi_{nm}(\mathbf{r}')^*$
 with 
\begin{eqnarray}
\mathcal{G}_{{\rm mf},11}(n,i\omega_j)=-\frac{u_n^2}{i\omega_j-E_n}-\frac{v_n^2}{i\omega_j+E_n}\nonumber\\
\mathcal{G}_{{\rm mf},12}(n,i\omega_j)=-\frac{u_nv_n}{i\omega_j-E_n}+\frac{u_nv_n}{i\omega_j+E_n}.
 \end{eqnarray}
 Here, $\omega_j=(2j+1)\pi T$  with $j=0,\pm 1,\ldots$ ($k_B=1$) is a Fermi Matsubara frequency.
  Also, $\mathcal{G}_{{\rm mf},22}(\mathbf{r},\mathbf{r}',i\omega_j)=-\mathcal{G}_{{\rm mf},11}(\mathbf{r}',\mathbf{r},-i\omega_j)$
 and $\mathcal{G}_{{\rm mf},21}(\mathbf{r},\mathbf{r}',i\omega_j)=\mathcal{G}_{{\rm mf},12}(\mathbf{r},\mathbf{r}',i\omega_j)^*$. Performing
  the Matsubara sums yields 
   $S_2=T\sum_l[s_2(i\omega_l)-\int d^2r\mathbf{d}({\mathbf r},i\omega_l)^\dagger\mathbf{d}({\mathbf r},i\omega_l)/g]/2$
  with 
  \begin{equation}
s_2(i\omega_l)= \int d^2rd^2r'\mathbf{d}({\mathbf r},i\omega_l)^\dagger{\mathbf M}({\mathbf r},{\mathbf r}',i\omega_l)\mathbf{d}({\mathbf r}',i\omega_l).
\end{equation}
In the intra-shell regime, the important monopole  correlations are between  time-reversed states in the same shell.   Keeping only   terms 
coupling $(n,m,\uparrow)$ and $(n,-m,\downarrow)$ yields  
${\mathbf M}({\mathbf r},{\mathbf r}',i\omega_l)=\sum_{nm}|\phi_{nm}({\mathbf r})|^2{\mathbf M}(n,i\omega_l)|\phi_{nm}^*({\mathbf r}')|^2$
with the matrix elements
\begin{eqnarray}
M_{11}(n,i\omega_l)=\frac{u_n^4}{i\omega_l-2E_n}-\frac{v_n^4}{i\omega_l+2E_n}\nonumber\\
M_{12}(n,i\omega_j)=\frac{u_n^2v_n^2}{2E_n-i\omega_l}+\frac{u_n^2v_n^2}{2E_n+i\omega_l}.
\label{Mmatrixelements}
 \end{eqnarray}
Also $M_{21}({\mathbf r},{\mathbf r}',i\omega_l)=M_{12}({\mathbf r},{\mathbf r}',i\omega_l)^*$ and 
$M_{22}({\mathbf r},{\mathbf r}',i\omega_l)=M_{11}({\mathbf r}',{\mathbf r},-i\omega_l)^*$. 

For a circular symmetric solution $\theta({\mathbf r})=\theta(r)$, the eigenvalue equation  for  the Goldstone mode becomes
 \begin{gather}
\lambda\theta(r)= \int d^2r'{\mathbf A}_{22}({\mathbf r},{\mathbf r}',\omega)\theta(r')=\nonumber\\
 \sum_{nm} \frac{R_{nm}(r)^2}{2\pi}A_{22}(n,\omega)\int_0^\infty dr' r'R_{nm}(r')^2\theta(r').
 \label{GoldstoneIntegral}
 \end{gather}
 In the intra-shell regime, we can ignore the $n,m$-dependence of the integral in (\ref{GoldstoneIntegral}) which then immediately gives the eigenfunction 
 $\theta(r)\propto \sum_nA_{22}(n,\omega)\rho_n(r)$. Inserting this function into 
 the eigenvalue equation  then yields the eigenvalue 
 \begin{equation}
 \lambda=\sum_nA_{22}(n,\omega)\frac{\int_0^\infty dr'r'\rho_{n_F}(r')^2}{\int_0^\infty dr'r'\rho_{n_F}(r')}
 \end{equation}
 where we again have ignored the weak $n,m$-dependence of the spatial  integrals. The Goldstone mode is determined by  $\lambda=1/g$ which together 
 with (\ref{Gdef}) and (\ref{twobody}) yields  (\ref{GoldstoneMode}). The derivation of (\ref{HiggsMode}) for the Higgs mode is identical apart from 
 the substitution $A_{22}\rightarrow A_{11}$.

\end{document}